\documentclass[12pt]{iopart}

\usepackage{graphicx}
\usepackage{epsfig}
\usepackage{bm}
\usepackage{color}
\usepackage{float}
\usepackage{dcolumn}

\def\be{\begin{equation}}
\def\ee{\end{equation}}
\graphicspath{ {images/} }
\newcommand{\bld}[1]{\boldsymbol{#1}}

\def\chsq{$\chi^2$~}
\newcommand{\cond}{\mbox{\rule[-0.5em]{.1mm}{1.2em}}}

\begin{document}
\title[Error Estimates of Theoretical Models]{Error Estimates of Theoretical Models: a Guide}
\date{Update \today}

\author{ J. Dobaczewski,$^{1,2}$ W. Nazarewicz,$^{3,4,1}$ and  P.-G. Reinhard$^{5}$}
\address{$^1$Institute of Theoretical Physics, Faculty of Physics, University of Warsaw,
ul. Ho{\.z}a 69, 00-681 Warsaw, Poland}
\address{$^2$Department of Physics, PO Box 35 (YFL), FI-40014
University of Jyv{\"a}skyl{\"a}, Finland}
\address{$^3$Department of Physics and
Astronomy, University of Tennessee, Knoxville, Tennessee 37996-1200, USA}
\address{$^4$Physics Division, Oak Ridge National Laboratory, P.O. Box  2008,
Oak Ridge, Tennessee 37831-6373, USA}
\address{$^5$Institut f\"ur Theoretische Physik II, Universit\"at Erlangen-N\"urnberg,
Staudtstrasse 7, D-91058 Erlangen, Germany}

\begin{abstract}
This guide offers suggestions/insights on uncertainty quantification of nuclear structure models. We discuss a simple approach to statistical error estimates, strategies to assess systematic errors, and show how to uncover inter-dependencies by correlation analysis. The basic concepts are illustrated through simple examples. By providing theoretical error bars on predicted quantities and using statistical methods to study correlations between observables, theory
can significantly  enhance  the
feedback between experiment and nuclear modeling.
\end{abstract}
\maketitle
%


\section{Introduction}

{\it ``Remember that all models are wrong; the practical question is
  how wrong do they have to be to not be
  useful."}~\cite{(Box87)} This quote, by George E.P. Box, a statistician
and a pioneer in the areas of quality control and Bayesian inference,
well applies to the nuclear many-body problem. When it comes to
nuclear modeling, uncertainties abound. Indeed, the nuclear
interaction, strongly influenced by in-medium effects, is not
perfectly known, and the same can be said about many operators
associated with observables. In addition the many-body equations are
difficult to crack, which forces nuclear modelers to introduce
simplifying assumptions.

The need for uncertainty estimates in papers involving theoretical
calculations of physical quantities has been long recognized in the atomic-physics
community. The current situation has been well captured by an Editorial in
Physical Review A~\cite{PhysRevA}:
\begin{quote}
{\it It is all too often the case that the numerical results
are presented without uncertainty estimates. Authors sometimes say
that it is difficult to arrive at error estimates. Should this be
considered an adequate reason for omitting them? [\ldots] There is a
broad class of papers where estimates of theoretical uncertainties
can and should be made. Papers presenting the results of theoretical
calculations are expected to include uncertainty estimates for the
calculations whenever practicable, and especially under the following
circumstances: (1) If the authors claim high accuracy, or
improvements on the accuracy of previous work; (2) If the primary
motivation for the paper is to make comparisons with present or
future high precision experimental measurements. (3) If the primary
motivation is to provide interpolations or extrapolations of known
experimental measurements.}
\end{quote}
This demand applies equally well, if not even more so, in nuclear
theory, where we have not a well settled {\em ab-initio}
starting point at hand. We are dealing almost everywhere with
effective theories justified in terms of general arguments, but whose
parameters are basically unknown and often cannot be deduced from {\em
  ab-initio} modeling. Consequently, those parameters are often
determined by fits to empirical data. This immediately raises the
question of the predictive power of such phenomenologically adjusted
theories; hence, there is a need for error estimates of the
predicted values.

We have here particularly in mind effective interactions or energy
functionals for nuclear structure and dynamics. These are usually
fitted to large sets of experimental data. The technique of
least-squares fitting, involved in these adjustments, opens
immediately the door to the well developed strategies for error
estimates from statistical (or regression)
analysis~\cite{(Bevington),(Bra97a)}. This is the least we can do, and
should do, towards delivering error estimates.  Besides mere error
estimates, statistical analysis is a powerful instrument to acquire
deeper insights into models, e.g., by determining weakly and strongly
constrained parameters or correlations between different observables.

However, a purely statistical analysis is not sufficient, as it does
not cover what is called the systematic errors, that is, missing
aspects of the modeling. Thus a broad discussion of extrapolation
errors must also address systematic errors. Unfortunately, systematic
errors are just those which cannot be dealt with in systematic
manner. It takes a deep insight into the related exact and approximate
theories to have a presentiment of possible pitfalls in a fit.

In the following, we address all three aspects: error estimates by
statistical analysis, attempts to assess systematic errors, and
uncovering inter-dependencies by correlation analysis.
A number of examples will be provided.  Some of the questions addressed are:
\begin{enumerate}
\item
How to estimate systematic and statistical errors of calculated quantities?
\item
What is model's information content?
\item
How to validate and verify theoretical extrapolations?
\item
What data are crucial for better constraining current nuclear models?
\end{enumerate}
A secondary objective of this guide is to set the stage for    the upcoming Focus Issue of  Journal of Physics G on  ``Enhancing the interaction between nuclear experiment and theory through information and statistics," which will contain many excellent examples of uncertainty quantification in nuclear modeling.

We hope that these notes  will serve as a useful guide for nuclear theorists, especially those who have not yet  embarked on the
uncertainty-quantification  journey.   In this context, we strongly recommend   a recent essay by Saltelli and Funtowicz~\cite{(Sal13)} on modeling issues at the science/policy interface; we found their checklist to aid in the
responsible development and use of models particularly insightful. The proposed
seven-rule checklist helps understanding the different sources of uncertainty
and their relative importance~\cite{(Sal13)}:
\begin{description}
\item[Rule 1:] {\it Use models to clarify, not to obscure}.   Many-parameter descriptions can be used at  an interim
  stage on the way to understanding but a fit seldom provides an answer. Remember the quote by von Neumann: ``{\em With four parameters I can fit an elephant, and with five I can make him wiggle his trunk}"?. Models should explain and simplify, but not make a situation more confused.
\item[Rule 2:] {\it Adopt an ``assumption hunting" attitude.} Are model assumptions
expressly stated or implicit/hidden? What in the model  is assumed to be irrelevant?
\item[Rule 3:] {\it Detect pseudoscience.} Be sure that uncertainties
have not been twisted to provide  desired results.
\item[Rule 4:] {\it Find sensitive assumptions before they find you.} Carry out
technically sound sensitivity studies.
\item[Rule 5:] {\it Aim for transparency.} Fellow scientists should be
able to replicate the results of your analysis.
\item[Rule 6:] {\it Don't just ``Do the sums right," but ``Do the right
sums"} --  to address the relevant  uncertainties.
\item[Rule 7:] {\it Focus the analysis.} Sensitivity
analysis of a many-parameter system cannot be done by merely changing one parameter at a time.
\end{description}

These notes are structured as follows.
In Sec.~\ref{sec:generalerrors}, we discuss the notion
of statistical and systematic errors.
Section \ref{sec:chi2} summarizes the technique of least-squares
optimization  and shows how to estimate statistical  errors.
In Sec.~\ref{sec:systematic}, we discuss
systematic errors and employ two pedagogical models to illustrate basic concepts.
In Sec. \ref{sec:correl}, we come back to statistical analysis and
show how it allows us to acquire deeper insights into model's information content.
Section \ref{sec:examples} contains selected examples
from recent work.

\section{Systematic and statistical model errors}
\label{sec:generalerrors}

In general, there are no surefire prescriptions for assigning error
bars to theory.  Model uncertainties have various sources, some are
rooted in experimental errors, some in model deficiencies. As well
put by Albert  Tarantola~\cite{(Tar05)}:
\begin{quote}
{\it The predicted values cannot, in general, be identical to the observed
values for two reasons: measurement uncertainties and model
imperfections. These two very different sources of error generally
produce uncertainties with the same order of magnitude, because, due
to the continuous progress of scientific research, as soon as new
experimental methods are capable of decreasing the experimental
uncertainty, new theories and new models arise that allow us to
account for the observations more accurately.}
\end{quote}
While the mutual interaction   between  experiment and theory resulting in a mutual reduction of uncertainties better applies  to atomic theory, as  theoretical uncertainties usually dominate experimental ones in nuclear modeling, the positive feedback described in the above quotation constitutes the  situation we  all should be striving to.

{\it\underline{The statistical model error}} is usually  quantifiable for many
models. Namely, when the model is based on parameters that were
fitted to large datasets, the quality of that fit is an indicator of
the statistical uncertainty  of the model's predictions. The commonly
employed tool to estimate statistical errors is the regression
analysis. Section~\ref{sec:chi2} shows how  to estimate statistical
errors by means of weighted least squares.

{\it\underline{The systematic model error}} is due to imperfect modeling:
deficient parametrizations, wrong assumptions,  and missing physics
due to our lack of knowledge. Since in most cases the perfect (exact,
reference) model is not available, systematic errors are extremely
difficult to estimate. Especially in the context of huge
extrapolations, no perfect strategy exists to assess systematic
errors, as some model features that are unimportant in the known
regions may become amplified, or even dominant, in the new domains.
Some commonly used ways to assess systematic model errors are
discussed in Sec.~\ref{sec:systematic}.

In all optimization problems, the key element of the approach is the
appropriate definition of the so-called penalty function. This
function, which depends on model parameters, experimental data, and
most often also on pre-defined parameters
specified by the modelers, gives us a one-dimensional
measure of the quality of the fit. By definition, model
  parameters leading to a smaller value of the penalty function are
considered to be superior to those leading to a larger value. The
optimization process is thus reduced to a minimization of the penalty
function. One cannot underestimate, and one should never forget about,
a possible fundamental influence of the definition of the penalty
function on the results of the optimization process. Through this
definition, a researcher may indeed exert influence on the
modeling -- this effect is as ubiquitous and fundamental as
the influence of an observer on a quantum system investigated.

For an exact model, the task is reduced to the optimization problem,
in which model's parameters are determined by comparing predicted
observables with carefully selected set of data. For such a perfect
model, the systematic error is zero, and the total error is
statistical, that is, it is given predominantly by the quality of
the measurement of the key data determining the model. Moreover,
exact models are characterized by an independence, or a
weak dependence, of the final result on the definition of the penalty
function~\cite{(Toi08)}.


In practice, nuclear models are imperfect, as most effective
models are, and the total uncertainty is usually dominated by systematic effects. How useful is it, therefore, to compute statistical
uncertainties of an imperfect model? There are, in fact, several good reasons to
embark on such a task:
\begin{enumerate}
\item
Statistical analysis yields uncertainties of model parameters. In
particular, by studying statistical errors on parameters one can
assess whether the dataset used to constrain the model is adequate
(in terms of quality and quantity).

\item
Statistical analysis can be used to compare different mathematical
formulations/assumptions and benchmark different approaches. In this
case, it is essential to use the same set of fit observables.

\item
The covariance matrix of the parameters tells us whether adding more
data makes sense. If the dataset is sufficiently diverse (that is, it
allows us to probe all directions in the model's parameter space)
and  large, the model may become over-constrained, and the resulting
statistical uncertainties may become small. In such a case, by
inspecting the non-statistical behavior of residuals
(which are the differences between the observed
and the estimated values) one can assess sources of missing
model features leading to systematic errors.

\item
Statistical method allows one to estimate the maximum model's
accuracy on a class of observables. If a higher accuracy is required,
further model refinements are needed.

\item
By assessing statistical errors of extrapolated quantities, one can
make a statement whether a model carries any useful information
content in an unknown domain.

\item
By using  Bayesian inference, one can test model's adequacy as
additional data are added, or additional evidence is  acquired (for
some recent nuclear  examples, see,  e.g.,
Refs.~\cite{(Ste10),Hig13,Szpak13}).

\item
The covariance matrix of the parameters enables one to compute
correlations between various observables predicted within a model.
This is a very useful tool when making new predictions and guiding
future experiments \cite{(Rei13a)}.

\item
A comparison of propagated statistical errors with residuals can be
one of the most powerful indicators of presence of missing aspects of
the model~\cite{(Gao13),(Kor13)}.
\end{enumerate}

\section{Estimating statistical errors}
\label{sec:chi2}

Let us consider a model having $N_p$ parameters
$\bld{p}=(p_1,...,p_{N_p})$ that are fitted to $N_d$ measured
observables ${\cal O}_i$ ($i=1,...,N_d$).   The steps are: define a
penalty function, minimize it with respect to the parameters
$\bld{p}$, construct the covariance matrix of the parameters, and
then apply the covariance matrix to estimate errors of
predictions by associating them with uncertainties in the parameter values.
The commonly used  penalty function is the \chsq objective function, by which we begin.

\subsection{The \chsq function}

We define the \chsq function for the parameter fit
as~\cite{(Bevington),(Bra97a),(Tar05)}
\begin{equation}\label{chi2}
\chi^2(\bld{p})
=
\sum_{i=1}^{N_d}
\frac{\left(\mathcal{O}_i(\bld{p})
      -
      \mathcal{O}^\mathrm{exp}_i\right)^2}
     {\Delta\mathcal{O}_i^2} ,
\end{equation}
where $\mathcal{O}_i(\bld{p})$ stands for the calculated values,
$\mathcal{O}^\mathrm{exp}_i$ for experimental data, and
$\Delta\mathcal{O}_i$ for adopted errors.
(It is to be noted, that when dealing with observables that
change by orders of magnitude (yields, half-lives), one must use
$\log(\mathcal{O})$ rather than $\mathcal{O}$ in Eq.~(\ref{chi2}).)
The model is thus defined
not only by the equations that are used to calculate the predicted
observables (that is, mathematical formulation and assumed model
space), but also by the dataset
$\left\{\mathcal{O}^\mathrm{exp}_i,i=1...N_d\right\}$ and
adopted errors $\left\{\Delta\mathcal{O}_i,i=1...N_d\right\}$
used to determine the parameters.

The adopted errors are determined as follows. Each one is the sum of
three components:
\begin{equation}\label{eq:composeerrors}
{\Delta\mathcal{O}_i}^2
=({\Delta\mathcal{O}_i^{\mathrm{exp}}})^2+({\Delta\mathcal{O}_i^{\mathrm{num}}})^2+
({\Delta\mathcal{O}_i^{\mathrm{the}}})^2 .
\end{equation}
Since the set of fit observables is usually  divided into types (masses, radii, \dots),   errors are adopted for each  data type separately.
The experimental
error, $\Delta\mathcal{O}_i^{\mathrm{exp}}$, is whatever the
experimentalists or evaluators quote in their datasets. The numerical error associated with the chosen
computational approach, ${\Delta\mathcal{O}_i^{\mathrm{num}}}$, is
also generally small, but may not be such for models based on, e.g.,
basis expansion methods~\cite{(Fur12)}.  In those cases, the numerical
error needs to be estimated.  The remaining part,
${\Delta\mathcal{O}_i^{\mathrm{the}}}$, is the theoretical error
due to inherent deficiencies of the model.

A judicious choice of the adopted errors ${\Delta\mathcal{O}_i^{\mathrm{the}}}$ is the crucial ingredient to the \chsq method for model development.
In practice,  the residuals of predicted observables should be confined to the range defined by ${\Delta\mathcal{O}_i^{\mathrm{the}}}$.  If only statistical errors are present, the residuals are stochastically distributed.
Since nuclear models are not perfect, however, trends in the residuals will appear due to systematic errors.

If different types of observables are present in the dataset, adopted errors have to be defined for each type.
For example, typical nuclear fits use one value of ${\Delta\mathcal{O}_i^{\mathrm{the}}}$  for  binding energies, another one for  r.m.s. radii, and so forth.
Each ${\Delta\mathcal{O}_i^{\mathrm{the}}}$ carries the same dimension as the observable $\mathcal{O}_i(\bld{p})$ thus rendering each contribution to \chsq dimensionless. In this way, different types of observables are combined  into one penalty function. The inverse square root $W_i=1/\sqrt{\Delta\mathcal{O}_i}$ defines the relative weight, wherewith the observable $\mathcal{O}_i(\bld{p})$ enters the optimization problem. By changing the values of $W_i$, one can control the impact of a particular observable, or a type of observables,  on the resulting parametrization.

It needs hardly be said that a certain degree of arbitrariness in choosing the weights $W_i$ is inevitable, as they can
be set individually for every  observable. In some cases, the weights vary from datum to datum \cite{(Mol95),(Eks13)}, while in many cases
equal weights $W_i$ are chosen for all observables of a given type. Clearly, there is no ``obvious" choice here~\cite{(Toi08)}, and various optimization protocols (driven by physics strategies)  are possible. This ambiguity  is one of  the primary reasons for a proliferation of  parametrizations in nuclear structure theory.

Fortunately, there is one guiding principle that comes to the rescue.
Remember that in the case of statistical fluctuations there is a consistency
between the distribution of residuals and the adopted error. Namely, the rules of statistical analysis require that the total penalty function at the minimum should be  normalized to
$N_d-N_p$, i.e., the average $\chi^2({\bld{p}_0})$ per degree of freedom should be  one:
\begin{equation}\label{cnorm}
\frac{\chi^2({\bld{p}_0})}{N_d-N_p}  \longleftrightarrow  1.
\end{equation}
This condition provides an overall scale for the normalization of the penalty function at the minimum and removes some of the arbitrariness in choosing the weights.

Now, the basic idea is to tune the ${\Delta\mathcal{O}_i^{\mathrm{the}}}$ so that it is consistent with the distribution of residuals, even if this distribution is non-statistical.
The relative weights between the
types of observables can thus be determined by requiring that the average
\chsq for each type is
\begin{equation}\label{Onorm}
  \sum_{i\in\mathrm{typ}}
  \frac{\left(\mathcal{O}_i(\bld{p})
      -
      \mathcal{O}^\mathrm{exp}_i\right)^2}
     {\Delta\mathcal{O}_i^2}
  =
  N_\mathrm{typ}\frac{N_d-N_p}{N_d},
\end{equation}
where $N_\mathrm{typ}$ is the number of data points of a given type.

It is thus clear that the values
$\Delta\mathcal{O}_i\approx{\Delta\mathcal{O}_i^{\mathrm{the}}}$
obeying the normalization  condition (\ref{Onorm}) cannot be chosen from the onset, but have to be
determined iteratively during the optimization process.  In
practice, the conditions  (\ref{cnorm}) or (\ref{Onorm}) are  seldom fulfilled   exactly. For example, it often
happens  that one wants to study variations of a fit while
keeping the adopted errors fixed \cite{(Klu09)},  which inevitably
changes the normalization condition (\ref{cnorm}). In order to deal
with such situations, we introduce a global scale factor $s$
such that
\begin{equation}
 \chi^2_\mathrm{norm}({\bld{p}_0})
 =
 \frac{\chi^2({\bld{p}_0})}{s} = N_d-N_p.
\label{chi2min}
\end{equation}
This amounts to a global readjustment
$\Delta\mathcal{O}_i\longrightarrow\Delta\mathcal{O}_i\sqrt{s}$ which
establishes correct normalization for $\chi^2_\mathrm{norm}$, but
leaves the relative weights untouched.
If  experimental and numerical errors are small compared to theoretical uncertainties, i.e.,
$\Delta\mathcal{O}_i={\Delta\mathcal{O}_i^{\mathrm{the}}}$, assumption
(\ref{chi2min})  defines a trivial scale factor,  which does not impact the
minimum $\bld{p}_0$. In the following, we shall
carry through all expressions with the scale factor $s$. This means
that the standard rules of statistical analysis will apply to the
normalized $\chi^2_\mathrm{norm}$.

Assumption (\ref{chi2min}), through its triviality is, in fact, the
only one that does not depend on the  researcher's choice.
In this way, the normalization of  \chsq  at its
minimum value at $\bld{p}_0$ is fixed by definition.  This
implies that one deals here with a model that is fundamentally
inaccurate and cannot describe simultaneously all the data within the
experimental and numerical errors alone. As it has been noted previously,  in the case of
negligible experimental and numerical errors, the minimization of \chsq
 does not depend on the condition (\ref{chi2min}); hence,
the scale $s$ can be computed after  determining $\bld{p}_0$.

In principle, an auxiliary  scale factor $s_{\rm typ}$ can be introduced for each data type, if the adopted theoretical errors are being adjusted during the optimization process. This amounts to   a readjustment
$\Delta\mathcal{O}_i\longrightarrow\Delta\mathcal{O}_i\sqrt{s_{\rm typ}}$ for $i\in\mathrm{typ}$ in each optimization step. The situation is particularly simple if one assumes the same weights for each data type. In this case the value of $s_{\rm typ}$ in a given step can be obtained directly
from the condition (\ref{Onorm}).
Such an iterative adjustment, leading eventually to $s_{\rm typ}=1$,  is recommended if the researcher has no
intuition about the expected theoretical error, and/or experimental and numerical errors are not negligible (see Ref.~\cite{(Mol95)} for a practical realization of an iterative adjustment of ${\Delta\mathcal{O}^{\mathrm{the}}}$ using the maximum likelihood method.)
In practical situations, however, this is seldom  the case, and a global scaling (\ref{chi2min}) following the minimization is fully adequate.

Another extreme case is the one when experimental/numerical errors are
so large that they mask the theoretical error entirely, and thus one
can set ${\Delta\mathcal{O}^{\mathrm{the}}}\equiv0$.  This case
  is often encountered in curve fitting of experimental data.  In such
  a situation the value of $\chi^2/(N_d-N_p)$ provides a direct estimate of
  the quality of the fit.

\subsection{Optimization}

The optimum parametrization $\bld{p}_0$ is the one that minimizes
the penalty function, in particular, the
$\chi^2$ function, with the minimum value of $\chi^2_0=\chi^2(\bld{p}_0)$.

\subsubsection{Pre-optimization.}
\label{sec:2.2.2}

A global model optimization becomes very involved when several categories of
fit-observables are considered. Such optimization procedures are expensive,
as they require a large number of model evaluations. Consequently, it is
always useful for the global optimization to have preliminary estimates for
the parameter values and their errors. An efficient pre-optimization method,
particularly convenient if observables are linear functions of model
parameters, is the linear regression algorithm employing the singular value
decomposition (SVD). This method has been used in the context of mass
fits~\cite{(Ber05),(Toi08)}, single-particle energies~\cite{(Kor08)}, and
pre-optimization of novel functionals~\cite{(Sto10)}. The advantage of the
SVD approach is that it can provide an efficient and accurate assessment of
model's error pertaining to a selected category of observables. In addition, it
provides an estimate of the effective size of the model parameter
space~\cite{(Ber05)}.  Many least-square solvers included in   optimization packages contain  an SVD truncation of the parameter space.

\subsubsection{Reasonable domain of model parameters.}

\begin{figure}
\center
\includegraphics[width=0.7\textwidth]{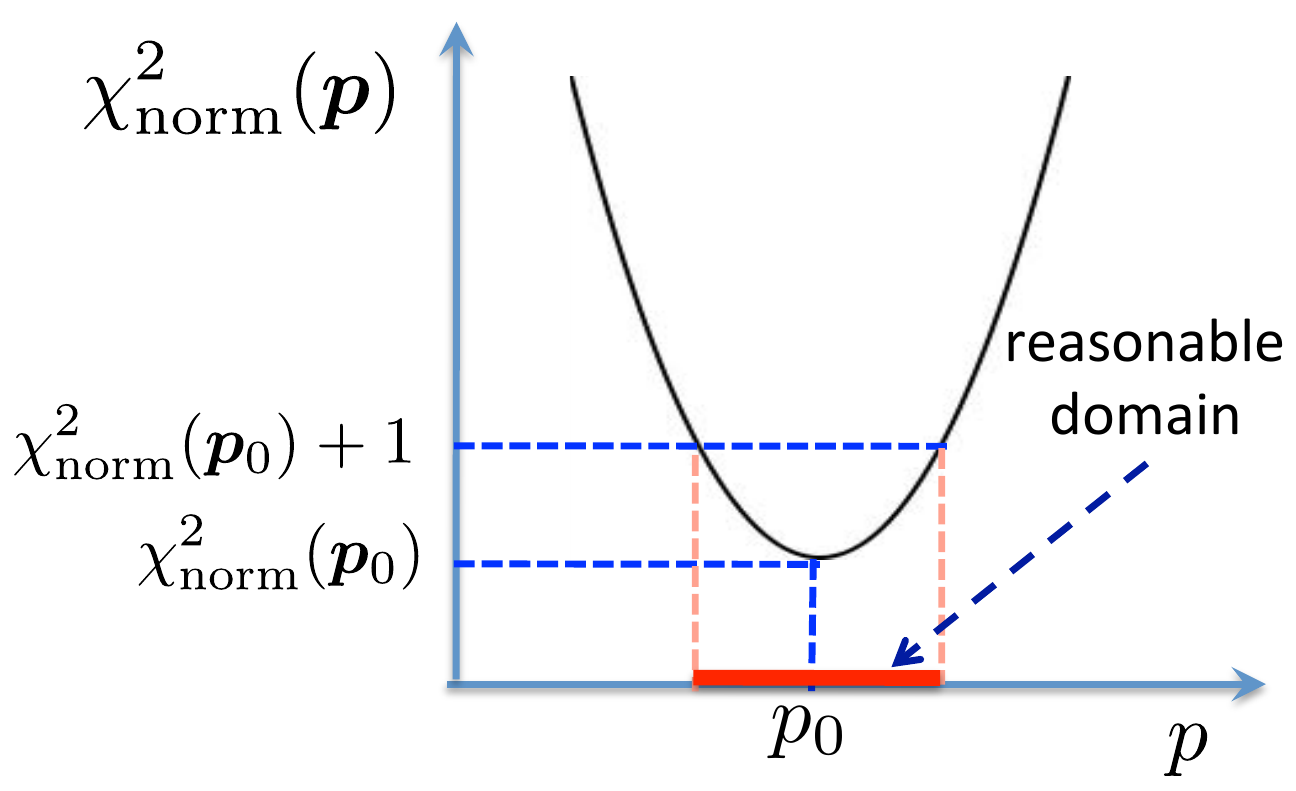}
\caption{\label{domain} The schematic illustration of a physically
reasonable domain around the $\chi_{\rm norm}^2$  minimum at
$\bld{p}=\bld{p}_0$.}
\end{figure}

Usually, most of the model space produces observables which are far from
reality. Therefore, one needs to confine the model space to a ``physically
reasonable'' domain around the minimum $\bld{p}_0$. Within this domain there
is a range of ``reasonable'' parametrizations $\bld{p}$ that can be
considered as delivering a decent fit, that is,
$\chi^2_\mathrm{norm}(\bld{p})\leq\chi^2_\mathrm{norm}(\bld{p}_0)+1$
(see Sec.~9.8 of Ref.~\cite{(Bra97a)}). As this range is usually rather
small, we can expand $\chi^2$ as
\begin{eqnarray}\label{chi2a}
  \chi^2(\bld{p})\!-\!\chi^2_\mathrm{0}
  &\approx&
  \sum_{\alpha,\beta=1}^{N_p} (p_\alpha-p_{0,\alpha})\mathcal{M}_{\alpha\beta}(p_\beta-p_{0,\beta}),
\\
  \mathcal{M}_{\alpha\beta}
  &=&
{\textstyle\frac{1}{2}}\partial_{p_\alpha}\partial_{p_\beta}\chi^2\cond_{\,\bld{p}_0},
\end{eqnarray}
that is, $\mathcal{M}$ is the Hessian matrix of \chsq at the minimum $\bld{p}_0$.
The reasonable parametrizations thus fill the confidence ellipsoid
given by
\begin{equation}\label{confidence}
  \frac{1}{s}(\bld{p}-\bld{p}_0)\hat{\mathcal{M}}(\bld{p}-\bld{p}_0)
  \leq 1,
\end{equation}
see Sec.~9.8 of Ref.~\cite{(Bra97a)} and Fig.~\ref{domain}.

\subsection{Statistical error}

Given a set of parameters  $\bld{p}$, any observable $A$ can be
within the model uniquely computed as $A=A(\bld{p})$. The value of
$A$ thus varies within the confidence ellipsoid, and this results in
some uncertainty $\Delta A$ of A.  Let us assume, for simplicity,
that the observable varies weakly with $\bld{p}$, such that one can
linearize it in the relevant range, that is,
\begin{equation}\label{linear}
A(\bld{p})\simeq
A_0+\bld{G}^A\cdot(\bld{p}- \bld{p}_0)
\quad\mbox{for}\quad
A_0=A(\bld{p}_0)
\quad\mbox{and}\quad
\bld{G}^A=\bm{\partial}_{\bm{p}}A\cond_{\,\bld{p}_0}.
\end{equation}
Let us,
furthermore, associate a weight $\propto\exp{\left(-\chi_{\rm norm}^2(\bld{p})\right)}$
with each parameter set~\cite{(Bra97a),(Tar05)}. The prescription for
assigning an error to the predicted value $A({\bf p_0})$ is the
following formula:
\begin{equation}\label{predicted_error}
 \overline{\Delta A^2}
  =
\sum_{\alpha\beta}G^A_\alpha\hat{\mathcal{C}}_{\alpha\beta}G^A_\beta,
\end{equation}
where $\hat{\mathcal{C}}$  is the covariance matrix.
Upon the assumption that the fitted observables $\mathcal{O}_i$ depend
only linearly on parameters $\bld{p}$, the covariance matrix
is simply related to the Hessian matrix as
\be\label{Mcovar}
\hat{\mathcal{C}} = s\hat{\mathcal{M}}^{-1} =
s\left (\hat{J}^T \hat{J}\right)^{-1},
\ee
where
\be\label{Jacobian} \hat{J}_{i\alpha} =
\frac{\partial_{p_\alpha}\mathcal{O}_i\cond_{\,\bld{p}_0}}{\Delta\mathcal{O}_i}
\ee
is the Jacobian matrix, which is inversely proportional to the
adopted errors. If the condition (\ref{chi2min}) is met with
  $s=1$, the expression for the covariance matrix simplifies to
$\left (\hat{J}^T \hat{J}\right)^{-1},$ see Table~\ref{tab:chis},  middle column. In particular, by taking $A(\bld{p}) = p_\alpha$, one obtains the expression for the statistical error of the model parameter $p_\alpha$:
$\Delta p_\alpha= \sqrt{s  \hat{\mathcal{C}}_{\alpha\alpha}}$.

We note that if the fitted observables weakly depend on some parameters,
the Hessian matrix becomes almost singular and the covariance
matrix becomes ill-conditioned. In such a case, errors or all
predicted values (\ref{predicted_error}) become unreasonably large. This shows again
that observables that weakly depend on model parameters
should not be fitted and parameters that have small impact on
the results should be removes from the model by proper pre-optimization
procedures, see Sec.~\ref{sec:2.2.2} and discussion in Sec. II.B of Ref.~\cite{(Fat11)}.

The Hessian ($\hat{\mathcal{M}}$), covariance ($\hat{\mathcal{C}}$), and Jacobian ($\hat{J}$) matrices
constitute the basic ingredients of the statistical-error analysis,
and thus should be routinely computed following the  optimization process.

\subsection{Dependence on the number of data points and on adopted errors}\label{datapoints}

It is worth noting that the Hessian matrix, $\hat{\mathcal{M}}=\hat{J}^T \hat{J}$,
increases linearly with the number $N_d$ of data
points constraining the model. This is best visible
in the case of identical observables $\mathcal{O}_i\equiv\mathcal{O}$ accompanied with
identical adopted errors $\Delta\mathcal{O}_i\equiv\Delta\mathcal{O}$,
whereupon one has
\be\label{identical}
\hat{\mathcal{M}}_{\alpha\beta}=\frac{N_d}{\Delta\mathcal{O}^2}
\left(\partial_{p_\alpha}\mathcal{O}\cond_{\,\bld{p}_0}\right)
\left(\partial_{p_\beta }\mathcal{O}\cond_{\,\bld{p}_0}\right) .
\ee
Therefore, the statistical uncertainty
(\ref{predicted_error}) does decrease with the number $N_d$ of data
points constraining the model. Indeed, if our model were exact
$({\Delta\mathcal{O}_i^{\mathrm{num}}}={\Delta\mathcal{O}_i^{\mathrm{the}}}=0)$, by
taking a very large number $N_d$ of fit observables, the model
observables would be determined with an arbitrary accuracy, that is,
\begin{equation}
\Delta{A}=\sqrt{\overline{\Delta A^2}}\sim
\frac{\Delta\mathcal{O}^{\mathrm{exp}}}{\sqrt{N_d}},
\end{equation}
and the precision would improve as the square root of the number of measurements
(data points), see Eq.~(4.12) of Ref.~\cite{(Bevington)}. In general,
as the number of uncorrelated data points  grows and the number of
parameters stays fixed, the confidence intervals become tighter. This
does not mean, of course, that by increasing $N_d$ we can make
predictions more accurate. At some point, adding more fit observables
makes little sense as the model error becomes dominated by the
systematic error, see Sec.~\ref{sec:systematic}.

Naturally, the precision of an exact model
also improves when all
experimental data are determined with smaller and smaller uncertainties
$\Delta\mathcal{O}^{\mathrm{exp}}$.
However, when the model
is inaccurate and the theoretical errors dominate, the normalization condition (\ref{chi2min})
applied to  observables of the same type, assuming identical weights,  gives:
\begin{equation}
\left(\Delta\mathcal{O}^{\mathrm{the}}\right)^2=\frac{N_d}{N_d-N_p}
\left(\mathcal{O}(\bld{p}_0)-\mathcal{O}^\mathrm{exp}\right)^2.
\end{equation}
That is, typical adopted theoretical errors are of the order of a
typical residual, and cannot be further decreased.

\section{Estimating systematic errors}
\label{sec:systematic}

A systematic error of a theoretical model is a consequence of missing
physics and/or poor modeling. Since in most cases the perfect model is
not available, systematic errors are very difficult to estimate. To
get some idea about systematic uncertainties, especially in the
context of extrapolations, one can adopt the following  strategies:
\begin{description}
\item[Analysis of residuals]
Study the distribution of residuals for a given
observable. For a perfect model, one should see a statistical
distribution. In most practical cases, however, one does see systematic
trends. These often allow us to guess the underlying missing
physics, see Sec.~\ref{sec:illust} for examples.
\item[Analysis of inter-model dependence]
Make a number of predictions ${\cal O}_j$ of an observable ${\cal O}$
using a set of $N_m$ reasonable and sufficiently different models
$M_j$ $j=1,..., N_m$ {\it well calibrated to existing data} and based
on different theoretical assumptions/optimization protocols.
Assuming that the biases introduced in different models are independent, one hopes that some randomization of systematic errors would take place.
The
predicted model-averaged value of an observable ${\cal O}$ can be
written as:
\begin{equation}
\overline{\cal O}_m={1\over N_m}\sum_{j=1}^{N_m} {\cal O}_j,
\end{equation}
with the corresponding systematic error
\begin{equation}
\Delta {\cal O}_{\rm syst, m}=\sqrt{{1\over N_m}\sum_{j=1}^{N_m} ({\cal O}_j-\overline{\cal O}_m)^2},
\end{equation}
which provides the scale of the model uncertainty.
\item[Comparison with existing data]
The systematic errors of fit observables can be
estimated by optimizing the model using a
large number data points to guarantee that the statistical error is
small. Then compute the r.m.s. deviation $\Delta {\cal O}_{\rm rms}$
from the known experimental data for a given type of observables
(e.g., masses or radii). The systematic error  of a predicted
observable ${\cal O}$ belonging to this type should be at least of
the order of $\Delta {\cal O}_{\rm rms}$.
\end{description}
 It is
recommended to combine the  strategies above by investigating both
$\Delta {\cal O}_{\rm syst, m}$ and $\Delta {\cal O}_{\rm rms}$.
Having estimated the systematic and statistical error
(\ref{predicted_error}), the predicted observable ${\cal O}$ can be
written as \begin{equation} {\cal O} = \overline{\cal O} \pm \Delta
{\cal O}_{\rm stat} \pm \Delta {\cal O}_{\rm syst}. \end{equation}
Let us emphasize again that the proposed analysis of adopted errors
(and hidden systematic errors) does not fully allow us to estimate the
contribution of the systematic error to extrapolations, as the
available data usually constrain only a limited region of the model
parameter space.

\subsection{Illustrative examples}
\label{sec:illust}

In this section, methodologies used to estimate statistical and
systematic errors are illustrated by means of schematic examples.
Those are followed in Sec.~\ref{sec:examples} by references to recent
studies aiming at uncertainty quantification in the context of
realistic nuclear modeling.

\subsubsection{Odd-Even Staggering Model.}
In this example, we illustrate the concept of statistical and
systematic errors using theoretically generated pseudo-data~\cite{(Ber13a)}:
\begin{equation}\label{OED}
\mathcal{O}^\mathrm{(exp)}_i = \delta (-1)^i + \varepsilon(i),~~~(i=1,2,... N_d),
\end{equation}
where $\delta$ stands for a magnitude of an odd-even staggering of a physical quantity (mass, radius,...), $\varepsilon(i)$ represents a white noise with zero mean (for both $i$-even and $i$-odd) and finite variance
\begin{equation}
\sigma^2={1\over N_d}\sum_{i=1}^{N_d}\varepsilon^2(i),
\end{equation}
and $N_d$ (an even number) is a number of data points.
It is assumed that $\delta \gg \sigma$ and $N_d \gg 1$.
To interpret the dataset (\ref{OED}) we employ two models
of one fitted observable:
\begin{description}
\item[Model A:]
$\mathcal{O}_{i=1}(\bld{p})=p_{\alpha=1}$ ($N_p=1$).
\item[Model B:]
$\mathcal{O}_{i=1}(\bld{p})=p_{\alpha=1} + p_{\alpha=2}(-1)^i$ ($N_p=2$).
\end{description}

\begin{figure}[htb]
\center
\includegraphics[width=0.7\textwidth]{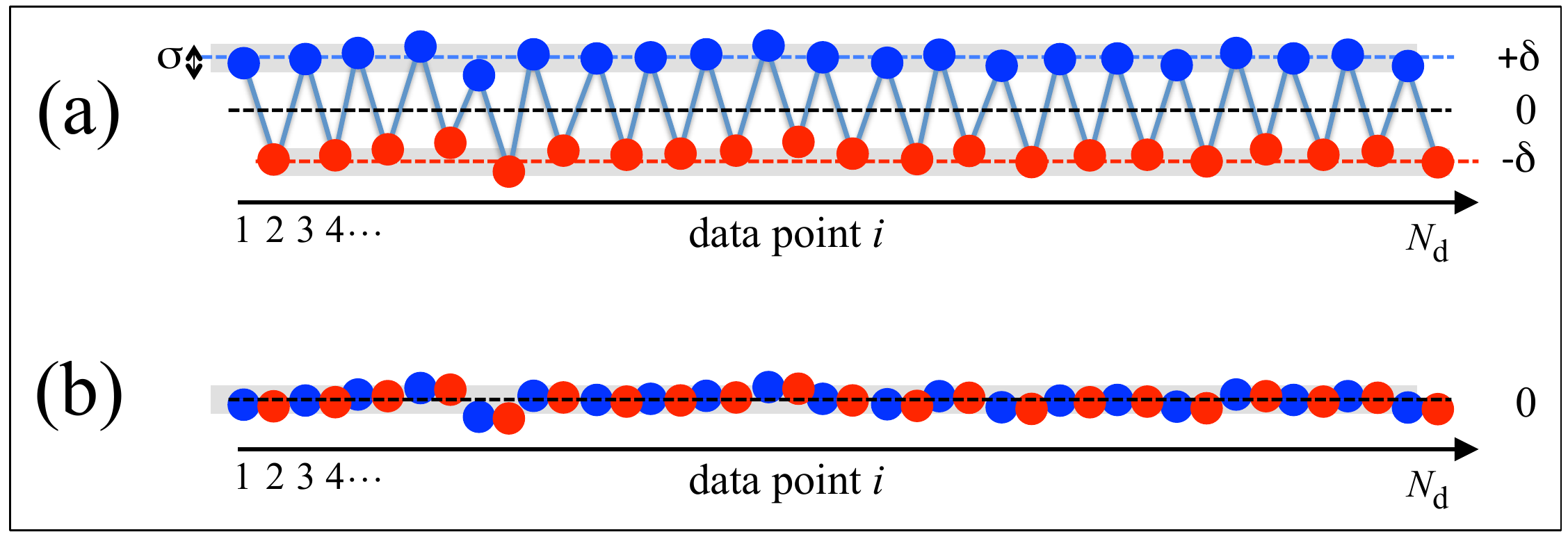}
\caption{\label{OES} Odd-even staggering residuals for Model A (top)
and Model B (bottom). The gray band indicates the variance of the
data noise.}
\end{figure}

Let us begin with Model A. It corresponds to a typical situation, in
which the nuclear model is imperfect, and experimental errors are
small. The minimization of $\chi^2$ yields $p_1=0$, and the condition
(\ref{chi2min}) yields ${\Delta\mathcal{O}_1} = {\Delta\mathcal{O}}
\approx \sqrt{\delta^2 + \sigma^2} \approx \delta$. The resulting
Jacobian matrix (\ref{Jacobian}) is $\hat{J}_{11}= 1/\delta$; hence,
the covariance matrix (\ref{Mcovar}) becomes
$(\hat{\mathcal{M}}^{-1})_{11}=\delta^2/N_d$. This is consistent with
the discussion in Sec.~\ref{datapoints}: the statistical uncertainty
(\ref{predicted_error})  decreases with the number $N_d$  of data
points constraining the model. The estimated statistical error on the
odd-even staggering, $\Delta {\cal O}_{\rm stat}=\delta/\sqrt{N_d}$,
can become very small if one takes very many data points. This of
course does not mean that our prediction is accurate. Indeed, by
inspecting the residuals of Model A shown in  Fig.~\ref{OES}(a), one
immediately concludes that their distribution is {\it not}
statistical. This suggests the presence of a large systematic error
that can be estimated as $\Delta {\cal O}_{\rm rms}=\delta$.
Consequently, for Model A, $\Delta {\cal O}_{\rm stat} \ll \Delta
{\cal O}_{\rm syst}$.

By inspecting the pattern of the residuals of Model A, one concludes
that this model is not adequate, and  this leads to a
two-parameter Model B. Here, the minimization of $\chi^2$ yields
$p_1=0, p_2=\delta$, and the condition (\ref{chi2min}) gives
${\Delta\mathcal{O}_1} = {\Delta\mathcal{O}} \approx \sigma$. The
resulting Jacobian matrix (\ref{Jacobian}) is $\hat{J}_{11}=
1/\sigma, \hat{J}_{12}= (-1)^i/\sigma$; hence, the covariance matrix
(\ref{Mcovar}) becomes $\hat{\mathcal{M}}^{-1}=\sigma^2/N_d \hat{I}$,
where $\hat{I}$ is a (2$\times$2) unity matrix. Figure~\ref{OES}(b)
shows the corresponding residuals: they are statistically distributed
around zero. This tells us that Model B is perfect, and its error is
statistical: $\Delta {\cal O}_{\rm stat} =\sigma$.

\subsubsection{Liquid Drop Model.}
In this example,  we test the $\chi^2$-optimization by using
theoretically-generated pseudo-data. To this end, $N_d=516$ even-even
nuclei with $6\le  Z\le 106$ listed in the Audi-Wapstra mass tables
were computed with the Skyrme functional SV-bas~\cite{(Klu09)} using the axial
HF+BCS approach.  Their binding energies were taken as pseudo-data to
which a four-parameter ($N_p=4$ ) LDM model for the total binding energy,
\begin{equation}\label{LDM}
 E(Z,N)
  =
  a_\mathrm{vol} A
  -
  a_\mathrm{surf}A^{-2/3}
  +
  a_\mathrm{sym}\frac{(N-Z)^2}{A}
  +
  a_\mathrm{C}\,Z^2A^{-1/3},
\end{equation}
was optimized. The adopted theoretical error on $E$ of 3.8 MeV was
tuned according to Eq.~(\ref{cnorm}).

\begin{table}[ht]
\caption{\label{LDMtable} Parameters of the LDM mass model
(\ref{LDM}) optimized to SV-bas masses (``fitted") compared to SV-bas
LDM constants obtained by means of the leptodermous expansion~\cite{(Rei06)}.
All values in MeV.}
\begin{center}
 \begin{tabular}{cccl}
    parameter &   SV-bas  &    fitted  & ${\cal O}_{\rm stat}$ \\
\hline
   $  a_\mathrm{vol}$ & -15.904  & -15.47  &    0.06 \\[-10pt]
   $a_\mathrm{surf}$  & 17.646   & 16.68   &   0.18  \\[-10pt]
   $a_\mathrm{sym}$   &  30.00   & 22.82   &  0.15 \\[-10pt]
   $a_\mathrm{C}$     &          &  0.702  &  0.004
 \end{tabular}
 \end{center}
 \end{table}

Table~\ref{LDMtable} compares the optimal parameters of the LDM model
(\ref{LDM}) with the  LDM parameters of SV-bas obtained by means of the
leptodermous expansion~\cite{(Rei06)}.  Due to the huge sample, one
obtains fairly  small statistical errors (see Sec.~\ref{datapoints}).  The deviation of the fit from
the   SV-bas LDM values  is much larger.  This  is
by no means surprising; to use the leptodermous expansion on bulk and surface
properties, one needs huge nuclei ($A\approx
1000-10000$) \cite{(Rei06)}. The
very incorrect  fitted symmetry energy demonstrates an
additional problem with the dataset used.  While it covers a
large range of mass numbers thus providing sufficient constraints on  isoscalar properties, the isospin range is  fairly limited, which results in  a poor determination of
isovector properties. The lesson learned from this exercise is that the dataset used is clearly inadequate:
the mass surface of known nuclei alone does not allow for a reasonable extraction of $a_\mathrm{sym}$~\cite{(Klu09)}.

\begin{figure}[htb]
\center
\includegraphics[width=0.7\textwidth]{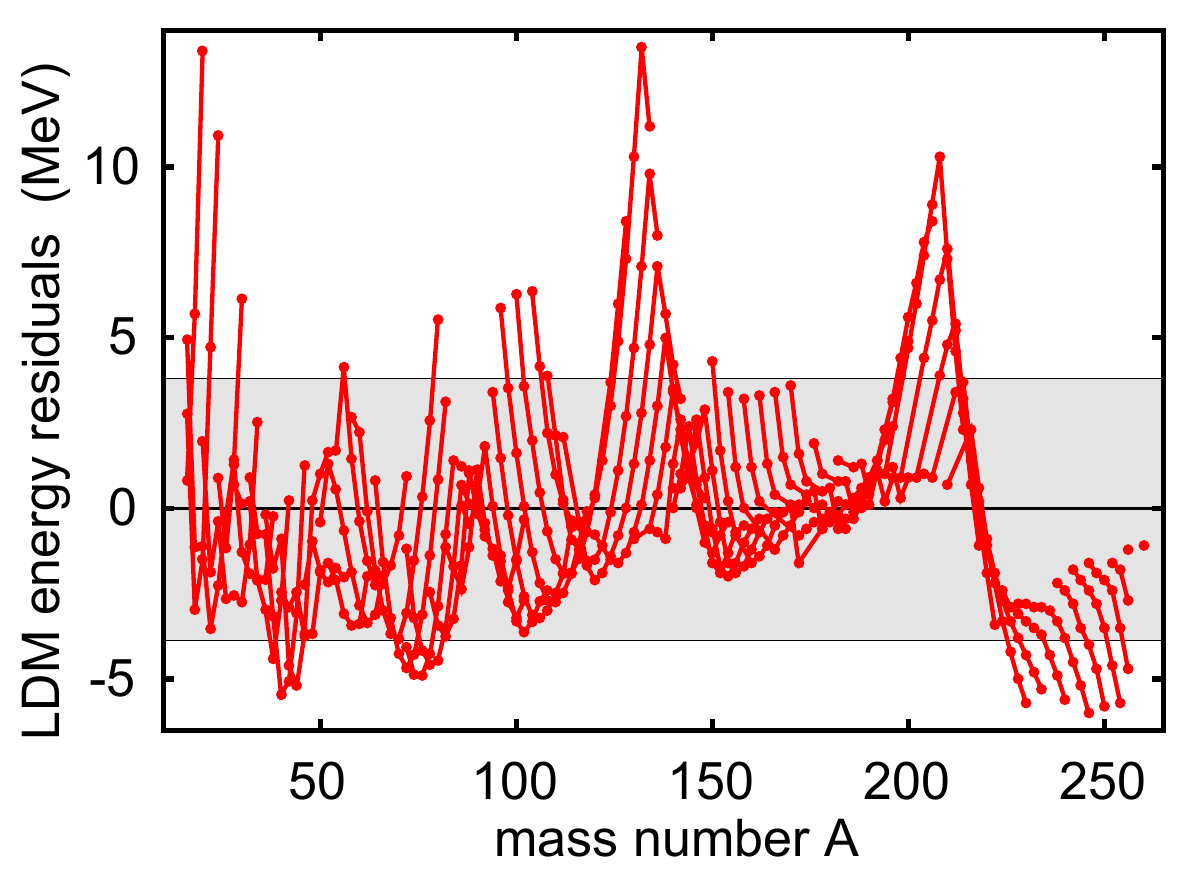}
\caption{\label{LDMfig} Total binding energy residuals obtained with
the LDM mass formula (\ref{LDM}). Isotopic chains are are connected
by  lines. The shaded area marks the r.m.s deviation from SV-bas
masses, $\pm 3.8$\,MeV.}
\end{figure}

Figure~\ref{LDMfig} shows the distribution of the resulting binding
energy residuals. The shaded area marks the band
corresponding to the final r.m.s error of 3.8\,MeV.  It is immediately seen that the binding energy residuals are not statistically distributed: there is a systematic trend due to the missing shell energy.  It is now clear, that the
error of the model (\ref{LDM}) on predicted masses is dominated by the
systematic component, which is at least 3.8\,MeV.  Of course, it is
well known that  systematic behavior of energy residuals can be partly  cured by adding shell corrections, and this makes the model quantitative (see, e.g., Fig.~1 of Ref.~\cite{(Mol95)}).

\section{Correlation analysis}
\label{sec:correl}

In this section, we come back to the rich  information
contained in the least-squares fits that provides
worthwhile insights into the actual (imperfect) model. We discuss here two aspects:
correlations between predicted observables
and probing the sensitivity of model parameters.

\subsection{Covariances}

A weighted average over the parameter space yields
the covariance between two observables $\hat{A}$ and $\hat{B}$, which
represents their combined uncertainty:
\begin{equation}\label{cova}
  \overline{\Delta A\,\Delta B}
  =
  \sum_{\alpha\beta}G_\alpha^A\hat{\mathcal{C}}_{\alpha\beta}G_\beta^B.
\end{equation}
For $A$=$B$, Eq.~(\ref{cova}) gives the variance $\overline{\Delta^2
A}$ that defines a statistical  uncertainty (error) of an observable (\ref{predicted_error}).
In addition, one can introduce a useful dimensionless
product-moment correlation coefficient~\cite{(Bra97a)}:
\begin{equation}
  {c}_{AB}
  =
  \frac{|\overline{\Delta A\,\Delta B}|}
       {\sqrt{\overline{\Delta A^2}\;\overline{\Delta B^2}}}.
\label{correlator}
\end{equation}
A value ${c}_{AB}=1$ means fully correlated and ${c}_{AB}=0$ --
uncorrelated. Variance, covariance, and the correlation coefficient are
useful quantities that allow us to estimate the impact of an observable on the
model and its parametrization.

\begin{figure}[htb]
\center
\includegraphics[width=0.7\textwidth]{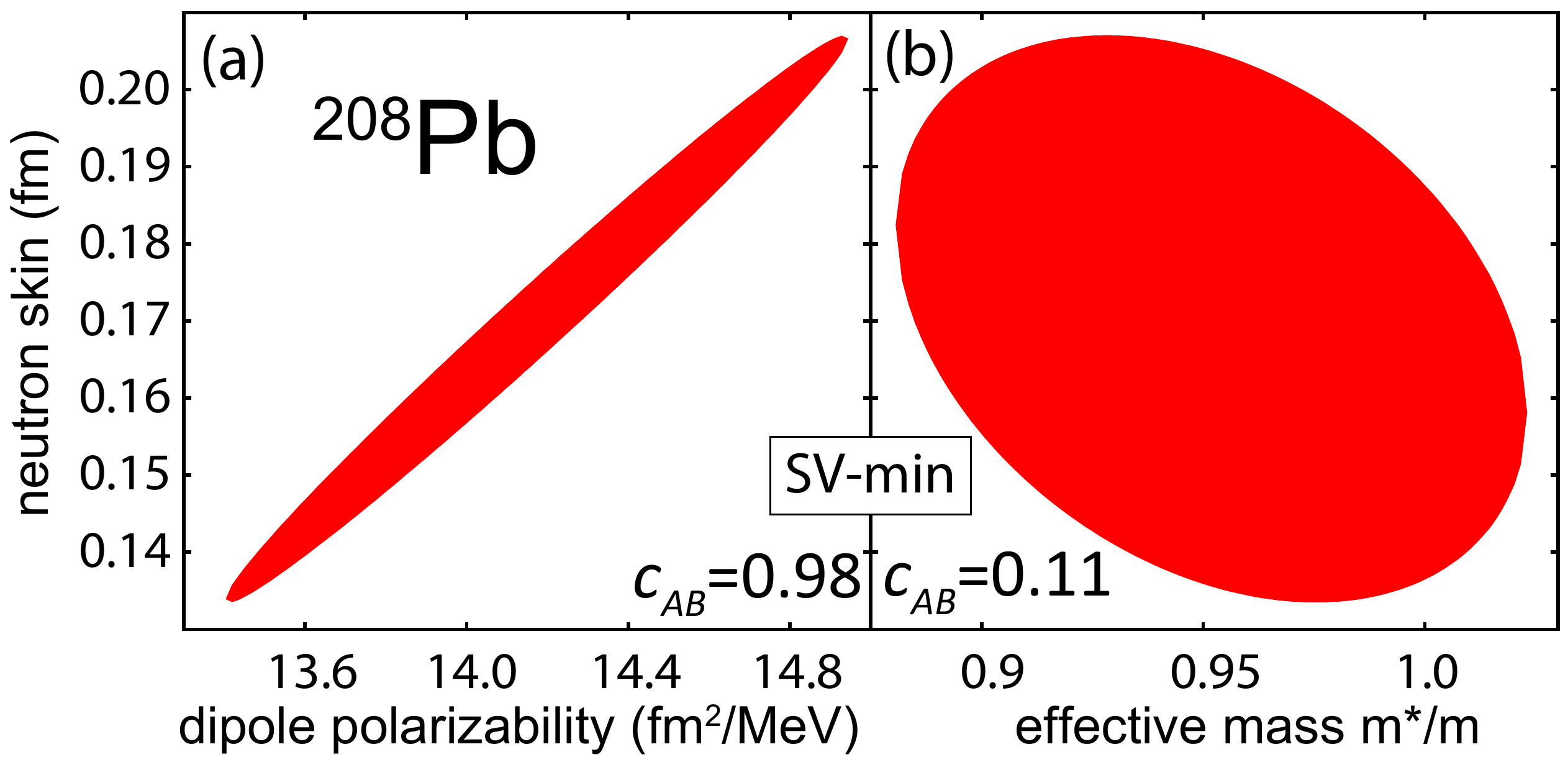}
\caption[T]{\label{fig:varellips-demo}
The covariance ellipsoids  for two  pairs of observables as
indicated. The filled area shows the region of reasonable domain
$\mathbf{p}$. Left: neutron skin and isovector dipole polarizability
in $^{208}$Pb. Right: neutron skin in $^{208}$Pb and  effective
nucleon mass $m^*/m$ in symmetric nuclear matter. (From
Ref.~\cite{Rei10}.)}
\end{figure}

Since the number of parameters is usually much smaller than the number of
observables, there must exist correlations between computed quantities.
Moreover, since the model space has been optimized to a limited set (and type) of observables, there must also exist correlations between model parameters.  Figure
\ref{fig:varellips-demo} shows covariance ellipsoids for two pairs of
observables in $^{208}$Pb that nicely illustrate the cases of strong
(neutron skin and isovector dipole polarizability; $c_{AB}$=0.98) and weak
(neutron skin and  effective
nucleon mass $m^*/m$ in symmetric nuclear matter; $c_{AB}$=0.11) correlation.

\begin{figure}
\center
\includegraphics[width=0.5\textwidth]{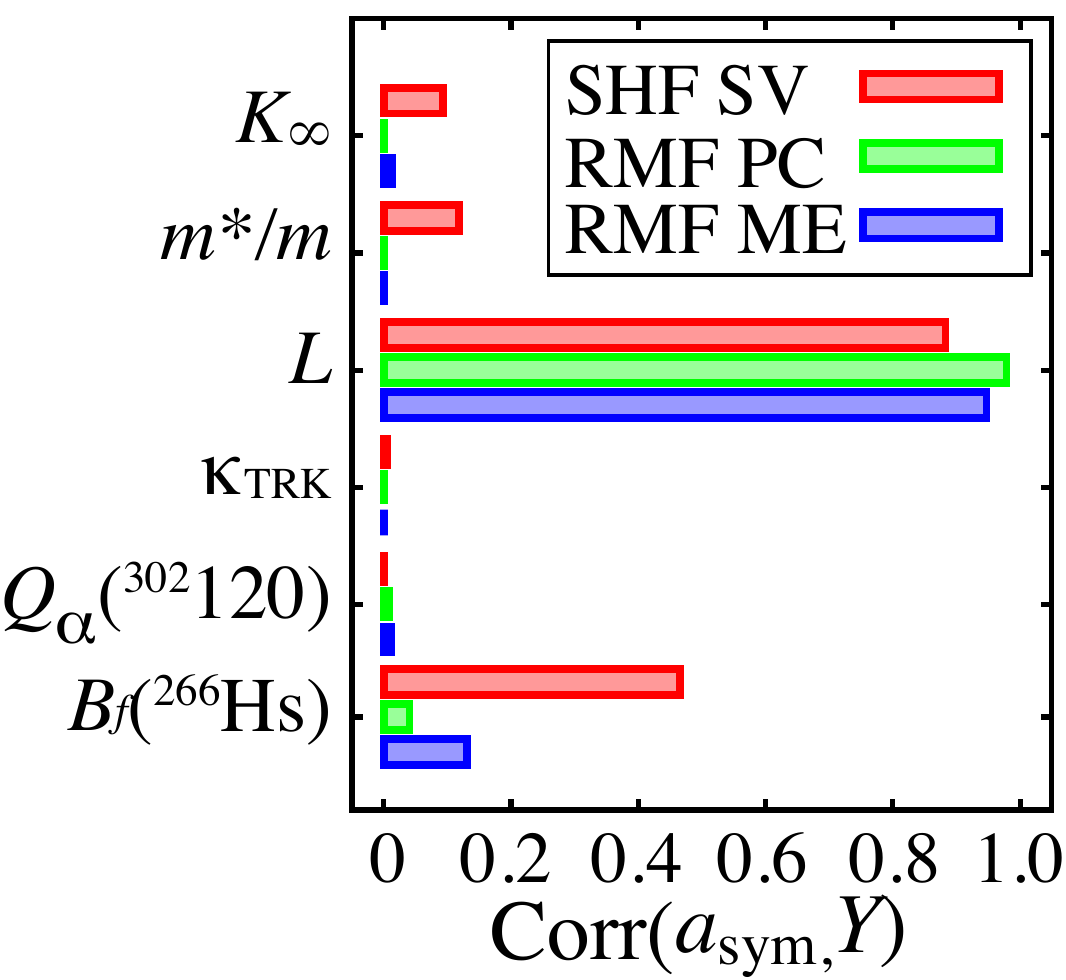}
\caption{\label{fig:symo-correl} The correlation (\ref{correlator})
between symmetry energy  and selected observables $Y$ (incompressibility
$K_\infty$, isoscalar effective mass $m^*/m$,  symmetry energy slope $L$,
TRK enhancement  $\kappa_\mathrm{TRK}$, $\alpha$-decay energy  $Q_\alpha$ in $^{302}120$, and  fission barrier $B_f$ in $^{266}$Hs)
computed in three models: SHF-SV, RMF-PC and RMF-ME. Results correspond to unconstrained
optimization employing the same strategy in all three cases.  (From Ref.~\cite{(Naz13)}.)
  }
\end{figure}

An example of the correlation analysis for the symmetry energy
$a_\mathrm{sym}$ is given in Fig.~\ref{fig:symo-correl} taken from the
survey, which compared predictions of Skyrme-Hartree-Fock (SHF) and
Relativistic-Mean-Field (RMF) energy density functionals (EDFs)~\cite{(Naz13)}.
The first four entries concern the basic  nuclear matter properties.
It is only for $L$, the density dependence of symmetry energy, that a
strong correlation with $a_\mathrm{sym}$ is seen.  The remaining two
entries  are $\alpha$-decay energy in yet-to-be-measured super-heavy
nucleus $Z=120, N=182$ and the fission barrier in $^{266}$Hs. The
data on $Z=120, N=182$ consistently do not correlate with
$a_\mathrm{sym}$. The correlation with fission barrier in $^{266}$Hs
exhibits an appreciable model dependence with some correlation in SHF
and practically none in RMF.

\subsection{Post-optimization: sensitivity tests}

\begin{figure}[ht]
\center
\includegraphics[width=0.7\textwidth]{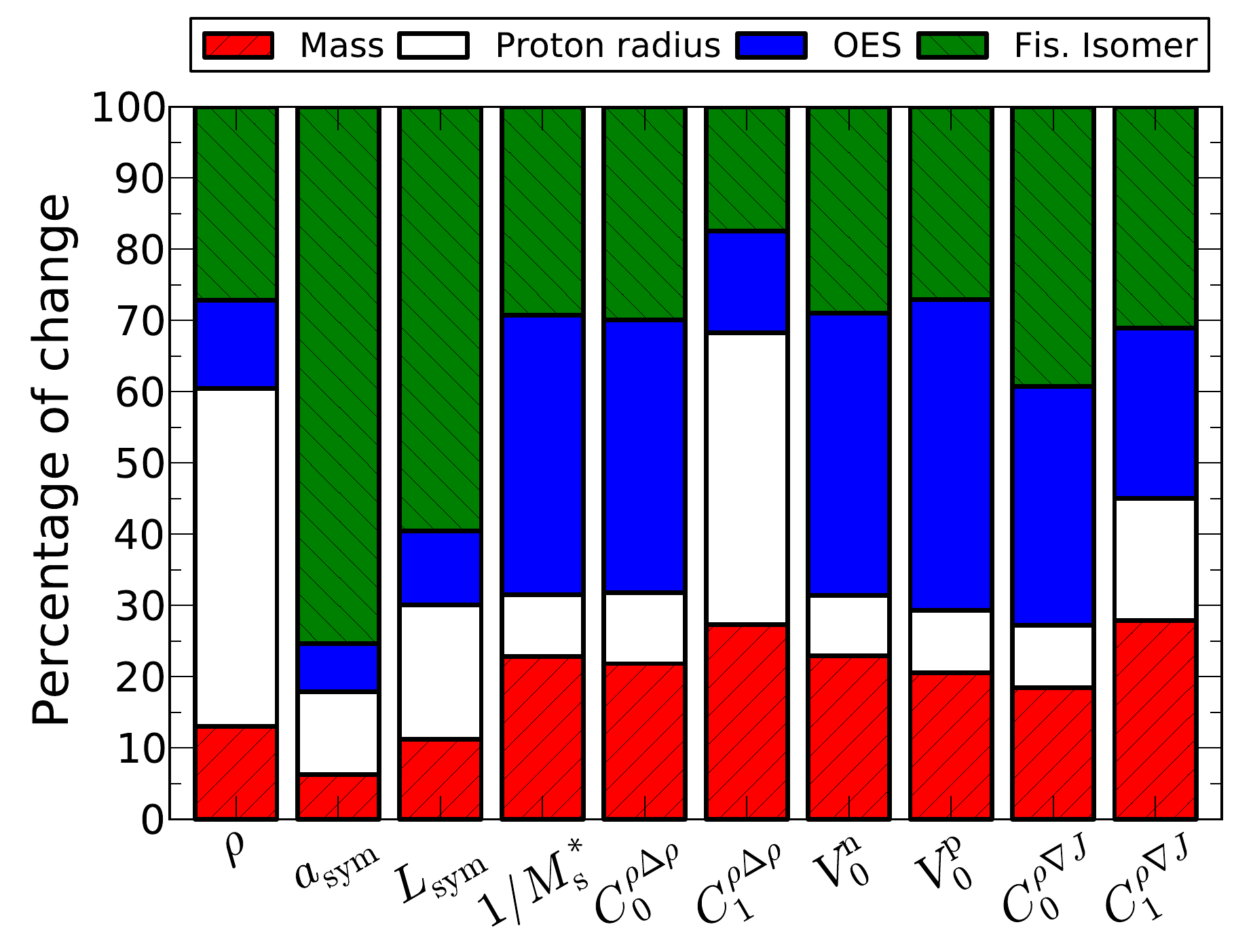}
\caption{Sensitivity of UNEDF1 energy density functional
parametrization  to different types of data entering the $\chi^{2}$
function. (From Ref.~\cite{(Kor12)}.)}
\label{fig:partsum}
\end{figure}

As discussed in Refs.~\cite{(Kor10),(Kor12)}, it is useful to study
the overall impact of each data type in the $\chi^{2}$
function on the obtained parameter set $\bld{p}_0$.
To this end, one can   employ the $N_p \times N_d$ sensitivity matrix
\begin{equation}
S(\bld{p})=\left[ \hat{J}(\bld{p})\hat{J}^{\rm T}(\bld{p})) \right]^{-1}
\hat{J}(\bld{p}).
\end{equation}
For each row in the sensitivity matrix (each parameter), one can
compute the partial sums over each  type of data. This gives us a
measure of the change of the parameter under a global change of all
the data of a given type. Figure~\ref{fig:partsum}, produced in the
context of UNEDF1 functional optimization ($N_p=10$, $N_d=115$),  shows the relative change
of parameter $p_\alpha$ when such an average datum of an observable
is changed.  The total strengths for each parameter were normalized
to 100\% and only relative strengths between various data types
(masses, proton radii, odd-even binding energy staggering, and fission isomer energies)
are shown.
A large percentage contribution from a given data type  tells that
$p_\alpha$ is very sensitive to changes in these data, and other data
types have little impact on it at the convergence point. Note that in
the example considered the  fission isomer excitation energies
represent less than 4\% of the total number of data points but
account for typically 30\% of the variation of the parameter set.
This kind of analysis, however, does not address the importance of an
individual datum on the optimal solution.

\begin{figure}[ht]
\center
\includegraphics[width=0.7\textwidth]{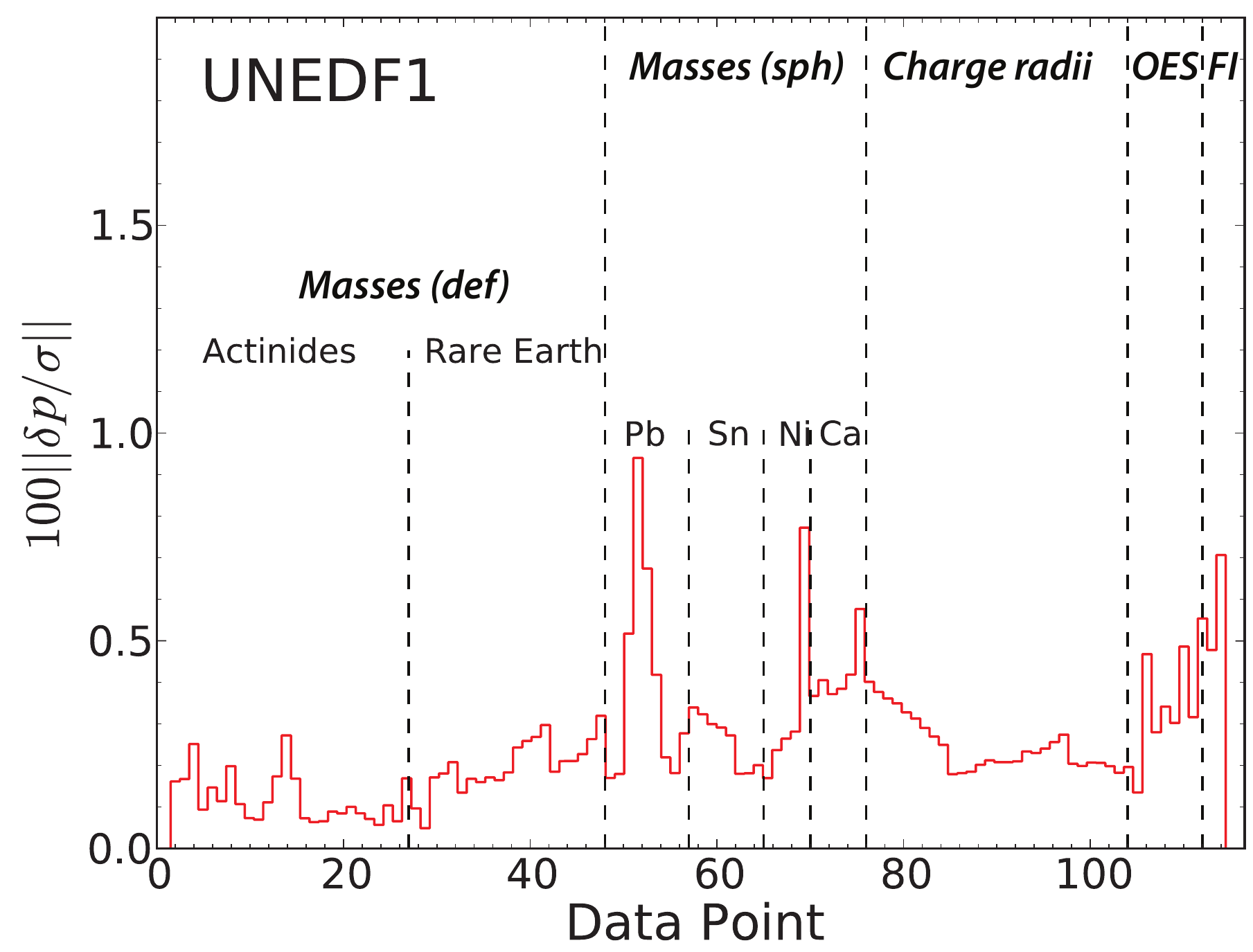}
\caption{Overall change in $\bld{p}$ for the UNEDF1 when the datum
$\mathcal{O}^\mathrm{(exp)}_i$  is changed by an amount of $0.1
{\Delta\mathcal{O}_i}$ one by one. The four rightmost data points
marked FI correspond to excitation energies of fission isomers.
(From Ref.~\cite{(Kor12)}.)}
\label{fig:sensitivity_UNEDF1}
\end{figure}

A  way to study the impact of an individual datum on the obtained
parameter set is  presented in Fig.
\ref{fig:sensitivity_UNEDF1} that shows the amount of
variation
\begin{equation}
\vert\vert {\delta\bld{p}}/{\sigma} \vert\vert
= \sqrt{\sum_{\alpha}\left(\frac{\delta p_{\alpha}}{\Delta p_\alpha}\right)^{2}}
\end{equation}
for the optimal parameter set when data points
$\mathcal{O}^\mathrm{(exp)}_i$ are changed by an amount of $0.1
{\Delta\mathcal{O}_i}$ one by one. As can be seen, the variations are
small overall, assuring us that the dataset was chosen correctly.
The masses of the double magic nuclei $^{208}$Pb and $^{58}$Ni seem
to have the biggest relative impact on the optimal parameter set. One
can also see that the sensitivity of the parameters on the new
fission isomer  data is larger than the average datum point. By
contrast, the dependence of the parametrization on the masses of
deformed actinides and rare earth nuclei is weaker.

\section{Examples of recent work}
\label{sec:examples}

In this section, we list examples of some recent theoretical work  involving
advanced optimization, error estimates, and covariance analysis.

\begin{figure}[ht]
\begin{center}
\includegraphics[width=0.7\textwidth,clip]{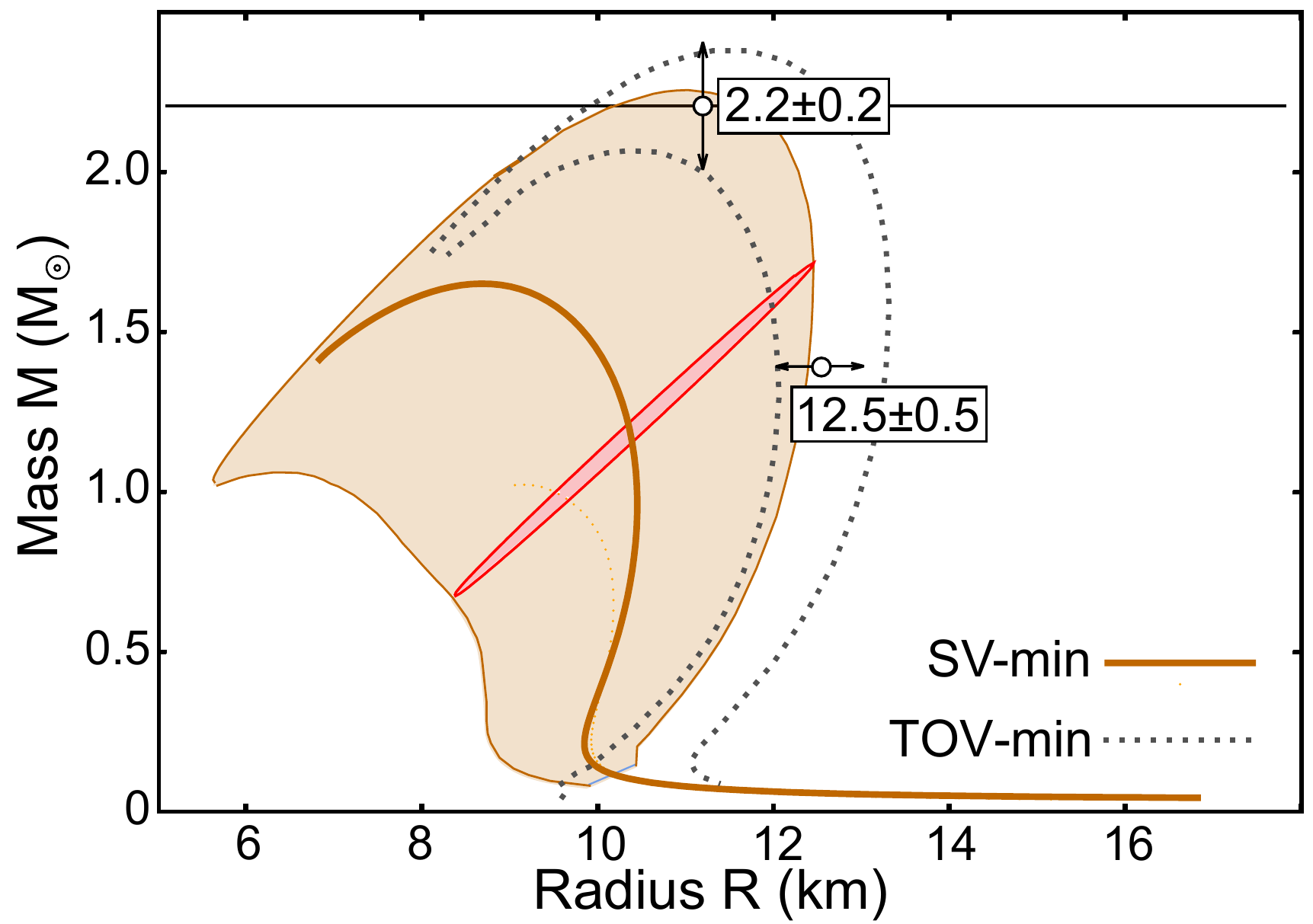}
\caption{
Mass-radius neutron star relation obtained for  SV-min and TOV-min
EDFs~\cite{(Erl13)}. The uncertainty band for SV-min is shown. This
band is estimated by calculating the covariance ellipsoid for the
mass $M$ and the radius $R$ at each point of the SV-min curve as
indicated by the ellipsoid. Also depicted (dotted lines) are
uncertainty limits for the newly developed functional TOV-min
optimized using exactly the same protocol as in earlier studies
pertaining to nuclei but now including neutron star data.}
\label{fig:uncert-SVmin}
\end{center}
\end{figure}

Optimization of nuclear energy density functionals (EDFs) using
different data categories were carried out for
nuclei~\cite{(Klu09),(Kor10),(Kor12),(Fat11)} and for nuclei and neutron
stars~\cite{(Erl13)}. Figure~\ref{fig:uncert-SVmin} shows the
predicted mass-radius relation of neutron stars for
SV-min~\cite{(Klu09)} and TOV-min~\cite{(Erl13)}  EDFs. In addition, the
estimated statistical uncertainty band for a prediction using SV-min
is shown. As both observables are correlated, it is not possible to
estimate the uncertainty for mass and radius separately. For this
reason, the error band is obtained by calculating the covariance
ellipsoid for each point of the $M(R)$ curve as indicated  in Fig.~\ref{fig:uncert-SVmin}.  The area covered by all covariance
ellipsoids can be viewed as the error band for a prediction
using SV-min. Based on this exercise, we can conclude that the
low-density part of the neutron matter equation of state, as given by
the commonly used nuclear EDFs optimized around the saturation
density {\it carries no information} on the high-density region.
Therefore, EDFs optimized to nuclear ground-state data cannot be used to predict, e.g., maximum mass of the neutron star; scrutinizing existing
functionals with respect to this quantity makes little sense. This
example nicely illustrates the point made earlier that by assessing
statistical errors of extrapolated quantities, one can make a
statement whether a model carries any useful information content in
an unknown domain. Here it does not.

The first solid attempt to carry out optimization of the nucleon-nucleon
interaction from chiral effective field theory at
next-to-next-to-leading order was done in Refs.~\cite{(Eks13),(Nav13),(Nav14)}.
In the phase-shift analysis of Ref.~\cite{(Eks13)}, it was assumed that the weights $W_i$ corresponding to phase shifts $\delta(q)$
decrease  with
the relative momentum $q$ -- to be consistent with the assumed order of the effective field theory. This is a good example of a situation, in which
physical arguments can impact  a  form of the penalty function.
The covariance analysis for the chiral constants in Refs.~\cite{(Nav13),(Nav14)}
enabled the uncertainty quantification of the interaction, parameter correlation, and predictions with error bars for deutron static properties.

In a number of recent papers, propagation of statistical
uncertainties in EDF models for separation energies and drip
lines~\cite{(Erl12)}, radii~\cite{(Kor13)},  and various structural
properties~\cite{(Gao13)} was carried out. A correlation testing analysis can be
found in Ref.~\cite{(Ber13)}. The recent papers
\cite{(Kor10),(Kor12),(Kor13)} contain sensitivity analysis and
statistical error budget for various observables.
Of particular importance to fission studies was a development of the
UNEDF1 parametrization \cite{(Kor12)} that is suitable for studies of strongly elongated nuclei. A
sensitivity analysis carried out for UNEDF1 has revealed the
importance of states at large deformations in driving the final fit.

\begin{figure}[ht]
\begin{center}
\includegraphics[width=0.9\textwidth,clip]{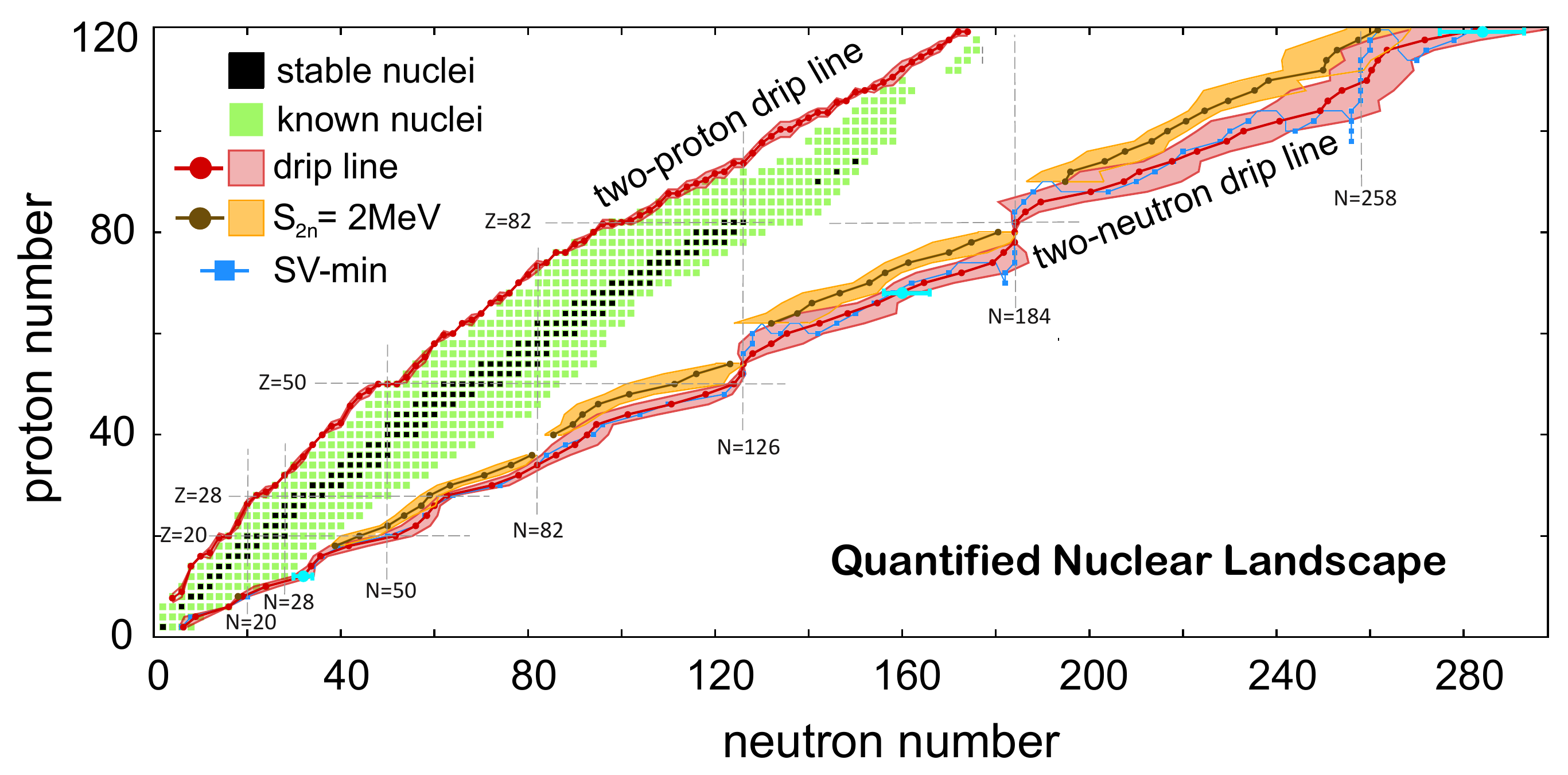}
\caption{Map of bound even-even nuclei as a function of the proton
number $Z$ and the neutron number $N$~\cite{(Erl12)}. Mean drip lines
and their uncertainties (red) were obtained by averaging the results
of different Skyrme-EDF models. (For a similar analysis using
covariant-EDF models, see Ref.~\cite{(Afa13)}.) The two-neutron drip
line of SV-min (blue) is shown together with the statistical
uncertainties at Z=12, 68, and 120 (blue error bars). The
$S_{2n}=2$\,MeV line is also shown (brown) together with its
systematic uncertainty (orange).}
\label{landscape}
\end{center}
\end{figure}

There are some studies, involving  inter-model analysis,  of
correlations and statistical and systematic errors for
nucleon-nucleon potentials and few-body systems \cite{(Ama13)},
drip lines~\cite{(Erl12),(Afa13)} (see Fig.~\ref{landscape}), neutron skins and
dipole polarizability~\cite{(Pie12),(Rei13),(Fat13)}, nucleon
densities~\cite{(Gao13)}, weak-charge form factor~\cite{(Rei13a)},
and neutron matter equation of state~\cite{(Ste10),(Ste13),Rei10,(Fat12),GandolfiNS,Heb13,Tews2013,(Erl13)}.

Examples and tests of statistical significance of the parameter
fitting procedures in the nuclear mean-field context using
phenomenological toy-models can be found in
Refs.~\cite{(Dud11),(Dud13),(Szpak11)}. Finally, an application of
the statistical likelihood analysis in the evaluation of fission
neutron data is described in Ref.~\cite{(Vog09)}.

\section{Summary}

{\it ``One important idea is that science is a means whereby learning is achieved, not by mere theoretical speculation on the one hand, nor by the undirected accumulation of practical facts on the other, but rather by a motivated iteration between theory and practice."}~\cite{(Box76)}
The essence of the scientific method is to explore the
positive feedback in the loop ``experiment-theory-experiment-..."
Based on experimental data, the theory is modified and
can be used to guide future measurements. The process is then
repeated, until the predictions are consistent with observations. The
process  can be enhanced if care is taken to
determine
parameter uncertainties and correlations, the errors of calculated observables, and the uniqueness and usefulness of an observable, that is,
its information content with respect to current theoretical models.

Since every nuclear model aiming at addressing the actual reality contains some parameters, the main source of statistical errors in nuclear modeling is
optimization of those parameters to experimental data. The good news is that various methods have been developed to assess model misfits.
Unfortunately, since {\it ``Essentially, all models are wrong"} \cite{(Box87)},  there is no perfect way to assess systematic uncertainties. One  option is to
use  high performance computing   to make
predictions using many models based on different assumptions. By means of an inter-model analysis
and by comparing with existing data, some information about
systematic errors can be deduced.

This guide and the references cited contain good  illustrations of  the seven rules of Ref.~\cite{(Sal13)}, quoted in the introduction,
when it comes to nuclear modeling. More  illuminating examples will soon appear in the   upcoming Focus Issue of  Journal of Physics G on  ``Enhancing the interaction between nuclear experiment and theory through information and statistics."

\bigskip
Numerous contributions from G.F. Bertsch are gratefully acknowledged.
We would like to thank A.N.  Andreyev, D. Higdon, J. Rosi{\'n}ski, W.
Loveland, and S. Wild for useful comments. This work was finalized
during the  Program INT-13-3 ``Quantitative Large Amplitude Shape
Dynamics: fission and heavy ion fusion" at the National Institute for
Nuclear Theory in Seattle; it was supported by the U.S. Department of
Energy under Contract No.  DE-FG02-96ER40963 (University of
Tennessee), No. DE-FG52-09NA29461 (the Stewardship Science Academic
Alliances program), No. DE-SC0008499    (NUCLEI SciDAC
Collaboration); by the Academy of Finland and University of
Jyv\"askyl\"a within the FIDIPRO programme; by the Polish National
Science Center under Contract No.\ 2012/07/B/ST2/03907; and by the
the Bundesministerium f\"ur Bildung und Forschung (BMBF) under
contract number 05P09RFFTB.
\bigskip

\appendix

\section{Different normalizations of $\chi^2$}
There are various forms of the least-squares function $\chi^2$ used in
the literature. Two of the most widely used options and their
consequences for Jacobian and variances are summarized in Table~\ref{tab:chis}.  The ``simple''
version (middle column) incorporates the basic assumption of properly
scaled $\chi^2$, namely $\chi^2(\mathbf{p}_0)\approx{N}_d$ and ignores
$N_p$ in the scaling assuming $N_d\gg N_p$, i.e., $s\approx 1$.

\begin{table}[H]
\caption{\label{tab:chis}
Different versions of the $\chi^2$ function and related quantities.
The scale factor $s$ is given by Eq.~(\ref{chi2min}): $s=\frac{\chi^2(\mathbf{p}_0)}{N_d-N_p}$.
}
\begin{center}
\begin{tabular}{l|c|c|c}
\hline
  quantity   &   \multicolumn{1}{|c|}{direct $\chi^2$}     &
\multicolumn{1}{|c|}{direct $\chi^2$, simple}
 & \multicolumn{1}{c}{normalized $\chi^2$} \\
\hline
 & & & \\[-15pt]
penalty function &
$ \sum_i\frac{\left(\mathcal{O}_i-\mathcal{O}_i^\mathrm{(exp)}\right)^2}
             {\Delta\mathcal{O}_i^2}$

&
$\sum_i\frac{\left(\mathcal{O}_i-\mathcal{O}_i^\mathrm{(exp)}\right)^2}
             {\Delta\mathcal{O}_i^2}$
&
$\frac{1}{N_d\!-\!N_p}
  \sum_i\frac{\left(\mathcal{O}_i-\mathcal{O}_i^\mathrm{(exp)}\right)^2}
             {\Delta\mathcal{O}_i^2}$
\\
Jacobian $J_{i\alpha}$
&
$
\frac{\partial\mathcal{O}_i}{\partial p_\alpha}\frac{1}{\Delta\mathcal{O}_i}
$
&
$
\frac{\partial\mathcal{O}_i}{\partial p_\alpha}\frac{1}{\Delta\mathcal{O}_i}
$
&
$
\frac{\partial\mathcal{O}_i}{\partial p_\alpha}\frac{1}{\Delta\mathcal{O}_i}
$
\\
variance $\Delta^2p_\alpha$
&
$s
 \left[\left(\hat{J}^\mathrm{T}\hat{J}\right)^{-1}\right]_{\alpha\alpha}
$
&
$
 \left[\left(\hat{J}^\mathrm{T}\hat{J}\right)^{-1}\right]_{\alpha\alpha}
$
&
$
 {\chi^2(\mathbf{p}_0)}
 \left[\left(\hat{J}^\mathrm{T}\hat{J}\right)^{-1}\right]_{\alpha\alpha}
$
\\
covariance $\hat{\mathcal{C}}$
&
$s  \left(\hat{J}^\mathrm{T}\hat{J}\right)^{-1}
$
&
$
  \left(\hat{J}^\mathrm{T}\hat{J}\right)^{-1}
$
&
$
 \chi^2(\mathbf{p}_0)
  \left(\hat{J}^\mathrm{T}\hat{J}\right)^{-1}
$
\\
\hline
\end{tabular}
\end{center}
\end{table}

\section*{References}
\bibliographystyle{unsrt}

\end{document}